\documentclass[twocolumn,floatfix,aps,nofootinbib,showpacs, showkeys]{revtex4}

\usepackage{bbold}
\usepackage[dvips]{epsfig}
\usepackage[english]{babel}
\usepackage{bbm}
\usepackage{verbatim}
\usepackage{array}
\usepackage{amsmath}
\usepackage{multirow}
\usepackage{amsfonts}
\usepackage{amssymb}
\usepackage{hyperref}
\usepackage{leading}
\usepackage[dvipsnames]{xcolor}
\hypersetup{colorlinks=true,linkbordercolor=Blue,linkcolor=Blue, citecolor=Blue}
\usepackage{subeqnarray}
\usepackage{setspace}
\usepackage{placeins}
\usepackage{float}
\usepackage{indentfirst} 

\usepackage{gensymb}

\def\I{\openone}
\def\openone{\mathbb I}

\begin{document}

\title{From Dipole spinors to a new class of mass dimension one fermions}

\author{R. J. Bueno Rogerio$^{1}$} \email{rodolforogerio@gmail.com}
\affiliation{$^{1}$Institute of Physics and Chemistry, Federal University of Itajub\'a , Itajub\'a, Minas Gerais, 37500-903, Brazil.}


\begin{abstract}
\noindent{\textbf{Abstract.}} In this letter, we investigate a quite recent new class of spin one-half fermions, namely \emph{Ahluwalia class-7 spinors}, endowed with mass dimensionality $1$ rather than $3/2$, being candidates to describe dark matter. Such spinors, under the Dirac adjoint structure, belongs to the Lounesto's class-6, namely dipole spinors. Up to our knowledge, dipole spinor fields have Weyl spinor fields as their most known representative, nonetheless, here we explore the \emph{dark} counterpart of the dipole spinors, which represents eigenspinors of the chirality operator.
\end{abstract}


\maketitle

\section{Introduction}\label{intro}
The Dirac and Majorana fields stand for a small part of a wide realm of the spinors fields. So it is natural to look for more possibilities to strengthen the pillars of Quantum Field Theory, and a method to accomplish that is extracting as much physical information relevant to the particles, and also to seek new candidates to explain physical phenomenon that are still open in physics --- as one of the most important open problem: dark matter.

Dark matter does not interact with the electromagnetic force, thus, it can not be directly experienced. Regardless of its evident gravitational effects \cite{elkograviton}, up to now, no associated \emph{dark} particle has ever been successfully detected, however, we have some strong candidates which may shed some light towards dark matter interpretation \cite{aaca,dharamnewfermions,tipo4epjc}.

The work deals with a particular new class of spinors which describe eigenspinors of chirality operator (chiral spinors), namely dipole spinors, in the lights of Lounesto's \cite{lounestolivro}. 
Weyl spinors, that are a particular class of dipole spinors with $U(1)$ gauge symmetry \cite{ferrari2017} were shown to belong to the Lounesto's class 6, being, thus, dipole spinor fields governed by the Weyl equation. However, the class 6 of dipole spinor fields further allocates mass dimension one spinors, whose dynamics, of course, is not ruled by the Weyl equation \cite{meert2018}. 

The unveiled Ahluwalia class-7 spinors stand for an entirely new class of spin one-half fermions endowed with mass dimensionality one, providing some new dark matter candidates \cite{dharamnewfermions}. Such a recently discovered fermions emerge from some (specific) set of arrangement of the Clifford algebra basis. The algorithm used to define these new spinors lies in an investigation of a linearly independent sets of the square roots of the $4\times 4$ identity matrix, the set of matrices $\Gamma_j$ \cite{schweber}. In addiction, we also highlight the quite new unexpected mass dimension three-half bosons of spin one-half \cite{dharamboson}.
We emphasize that the spinors at hands stand for a complete set of eigenspinors of $\Gamma_7$ and since $\Gamma_7$ commutes with the chirality operator, they are also eigenspinors of the chirality operator. 

According to Lounesto, dipole spinors are described by (massless) neutrinos \cite{lounestolivro}. Nonetheless, we show the possibility to \emph{enlarge} the dipole spinors understanding, since Ahluwalia class-7 spinors brings to light the possibility of obtaining dark and massive dipole spinors.
Generally speaking, mass dimension one fermions compose what is usually known as ``\emph{Beyond the Standard Model}'' of particle physics --- such as Elko spinors \cite{mdobook} and flag-dipole spinors \cite{tipo4epjc}. Interestingly enough, a possible path to deep understand and also look towards describing \emph{Beyond the Standard Model} physics is accomplished by exploring unexpected ideas, especially taking into account the actual status of theoretical physics.

As it can be directly seen, the Ahluwalia class-7 spinors opens up the possibility of providing a natural self-interacting dark matter candidate.
The dark feature encoded on these new spin one-half matter fields arises from the following observation: the mass dimension incompatibility with the usual fermions of the Standard Model of particle physics, moreover, as highlighted in \cite{dharamnewfermions}, Ahluwalia class-7 fermions cannot enter the standard model doublets --- standing for natural dark matter candidates with unsuppressed quartic self-interaction, with mass dimensionality found to be $1$ rather than $3/2$. Accordingly to the usual machinery, a form to describe charged matter fields is accomplished via the Dirac formalism, whereas the dark sector may be defined upon mass dimension one fermions.


The manuscript is organized as follows: in the next section, we explore the mathematical device which allow to define the new set of mass dimension one spinors. Going further, we search for the physical content encoded on the new set of dark dipole spinors. Finally, in Sect.\ref{remarks} we conclude.

\section{Outline and discussions}\label{outline}
The well known Lounesto's spinor classification stand for a comprehensive spinor categorization based on the bilinear covariants, disclosing the possibility of a large variety of spinors. Comprising regular and singular spinors which include the cases of Dirac, flag-dipole, Majorana and Weyl spinors \cite{bonorapandora}. 

The aforementioned classification stands for a geometrical classification and usually classify spinors according to their physical information, the criterion lies on the so-called bi-spinorial densities \cite{crawford1,crawford2}. 

The mass dimension one theory is still an open issue in QFT \cite{mdobook,tipo4epjc}. Then, we believe that a wide and interesting content still ``hidden'' in the mass dimension one theory. 

We start the discussions bringing to scene the new class of mass dimension one fermions recently developed in \cite{dharamnewfermions}. Such set of mass dimension one fermions arise from the direct computation of the eigenspinors of a specific set of matrices, recalling the well known linearly independent square roots of identity \cite{schweber}, given by
\begin{eqnarray}
&&\I  \\
&&i\gamma_1 \;\;\; i\gamma_2 \;\;\; i\gamma_3 \;\;\; \gamma_0 \\
&&i\gamma_2\gamma_3 \;\;\; i\gamma_3\gamma_1 \;\;\; i\gamma_1\gamma_2 \;\;\; \gamma_0\gamma_1 \;\;\; \gamma_0\gamma_2\;\;\; \gamma_0\gamma_3 \\
&&i\gamma_0\gamma_2\gamma_3 \;\;\; i\gamma_0\gamma_1\gamma_3 \;\;\; i\gamma_0\gamma_1\gamma_2 \;\;\; \gamma_1\gamma_2\gamma_3  \\
&&i\gamma_0\gamma_1\gamma_2\gamma_3 
\end{eqnarray}
denoting the above set of matrices by $\Gamma_j$, $j=1,\cdots, 16$, in which $\Gamma_1=\I$ and $\Gamma_{16}=i\gamma_0\gamma_1\gamma_2\gamma_3$. The factor $i$ ensure the following: $\Gamma^{2}_{j}=+\I$, in addition, ensuring real eigenvalues. For completeness, note that $\{\Gamma_{j},\Gamma_k\}=2\delta_{jk}\I$ for $j,k=2,\cdots,16$.
This representation is irreducible, any other representation can be expressed in terms of the elements above \cite{schweber}. Such a set of matrices may provide a range of possibilities to define new mass dimension one fermions and also mass dimension three-halves bosons \cite{dharamnewfermions,dharamboson}. 

The method developed in \cite{dharamnewfermions} lies on the complete set of eigenspinors provided by $\Gamma_{7}=i\gamma_3\gamma_1$. Thus, its rest spinors are
\begin{eqnarray}
\lambda_1(\textbf{0})= \left(\begin{array}{c}
0 \\ 
0 \\ 
-i \\ 
1
\end{array} \right),\quad \lambda_2(\textbf{0})= \left(\begin{array}{c}
0 \\ 
0 \\ 
i \\ 
1
\end{array} \right), 
\end{eqnarray}
and 
\begin{eqnarray}
\lambda_3(\textbf{0})= \left(\begin{array}{c}
-i \\ 
1 \\ 
0 \\ 
0
\end{array} \right),\quad \lambda_4(\textbf{0})= \left(\begin{array}{c}
i \\ 
1 \\ 
0 \\ 
0
\end{array} \right). 
\end{eqnarray}
Notice that $\lambda_1$ and $\lambda_3$ correspond to eigenvalue $+1$ and the remaining two spinors correspond to $-1$ eigenvalue. Henceforth, we must call these spinors \emph{Ahluwalia spinors of class-7}.
So far, we introduced the rest-frame spinors. To define the spinors for an arbitrary momentum, $\lambda_{j}(\textbf{p})=\kappa\lambda_{j}(\textbf{0})$, it suffices acting with the boost operator
\begin{eqnarray}
\kappa = \sqrt{\frac{E+m}{2m}}\left(\begin{array}{cc}
\I+ \frac{\vec{\sigma}\cdot\vec{\textbf{p}}}{E+m} & 0 \\ 
0 & \I- \frac{\vec{\sigma}\cdot\vec{\textbf{p}}}{E+m}
\end{array} \right),
\end{eqnarray} 
in which $\boldsymbol{\sigma}$ stands for the Pauli matrices. 
Thus, the $\lambda$'s spinors under the Dirac dual $\bar{\lambda}_{j}=\lambda^{\dag}_{j}\gamma_0$, furnish the following: $\sigma=\bar{\lambda}\lambda$, $\omega=i\bar{\lambda}\gamma_5\lambda$ and $S_{\mu\nu}=\frac{i}{2}\bar{\lambda}\gamma_{\mu}\gamma_{\nu}\lambda$ identically vanishing. And the set of non-vanishing bilinear quantities: $J_{\mu}=\bar{\lambda}\gamma_{\mu}\lambda$ and $K_{\mu}=-\bar{\lambda}\gamma_5\gamma_{\mu}\lambda$, standing for the conserved current and the the axial-vector current, respectively. Thus, the $\lambda_1$ spinor yield  
\begin{eqnarray}\label{bilinearsset1}
J_{0_{\lambda_1}} = -K_{0_{\lambda_1}} = \frac{(E+m)^2 - 2p_y(E+m)+p^2}{m(E+m)}, \nonumber
\end{eqnarray}
\begin{eqnarray}
J_{1_{\lambda_1}} = -K_{1_{\lambda_1}} = -\frac{2p_x(E+m-p_y)}{m(E+m)}, \nonumber
\end{eqnarray}
\begin{eqnarray}
J_{2_{\lambda_1}}=-K_{2_{\lambda_1}}=\frac{(E+m)^2 - 2p_y(E+m)-p_x^2+p^2_y-p^2_z}{m(E+m)}, \nonumber
\end{eqnarray}
\begin{eqnarray}
J_{3_{\lambda_1}}=-K_{3_{\lambda_1}}= -\frac{2p_z(E+m-p_y)}{m(E+m)}.
\end{eqnarray}
The $\lambda_2$ spinors provide
\begin{eqnarray}\label{bilinearsset2}
J_{0_{\lambda_2}} = -K_{0_{\lambda_2}} = \frac{(E+m)^2 + 2p_y(E+m)+p^2}{m(E+m)}, \nonumber
\end{eqnarray}
\begin{eqnarray}
J_{1_{\lambda_2}} = -K_{1_{\lambda_2}} = -\frac{2p_x(E+m+p_y)}{m(E+m)}, \nonumber
\end{eqnarray}
\begin{eqnarray}
J_{2_{\lambda_2}}=-K_{2_{\lambda_2}}=-\frac{(E+m)^2 + 2p_y(E+m)-p_x^2+p^2_y-p^2_z}{m(E+m)}, \nonumber
\end{eqnarray}
\begin{eqnarray}
J_{3_{\lambda_2}}=-K_{3_{\lambda_2}}= -\frac{2p_z(E+m+p_y)}{m(E+m)}.
\end{eqnarray}
The $\lambda_3$ furnish the following bilinear forms
\begin{eqnarray}\label{bilinearsset3}
J_{0_{\lambda_3}} = K_{0_{\lambda_3}} = \frac{(E+m)^2 + 2p_y(E+m)+p^2}{m(E+m)}, \nonumber
\end{eqnarray}
\begin{eqnarray}
J_{1_{\lambda_3}} = K_{1_{\lambda_3}} = -\frac{2p_x(E+m+p_y)}{m(E+m)}, \nonumber
\end{eqnarray}
\begin{eqnarray}
J_{2_{\lambda_3}}=K_{2_{\lambda_3}}=-\frac{(E+m)^2 + 2p_y(E+m)-p_x^2+p^2_y-p^2_z}{m(E+m)}, \nonumber
\end{eqnarray}
\begin{eqnarray}
J_{3_{\lambda_3}}=K_{3_{\lambda_3}}= -\frac{2p_z(E+m+p_y)}{m(E+m)}.
\end{eqnarray}
Finally, for $\lambda_4$ spinors we have
\begin{eqnarray}\label{bilinearsset4}
J_{0_{\lambda_4}} = K_{0_{\lambda_4}} = \frac{(E+m)^2 - 2p_y(E+m)+p^2}{m(E+m)}, \nonumber
\end{eqnarray}
\begin{eqnarray}
J_{1_{\lambda_4}} = K_{1_{\lambda_4}} = -\frac{2p_x(E+m-p_y)}{m(E+m)}, \nonumber
\end{eqnarray}
\begin{eqnarray}
J_{2_{\lambda_4}}=K_{2_{\lambda_4}}=-\frac{(E+m)^2 - 2p_y(E+m)-p_x^2+p^2_y-p^2_z}{m(E+m)}, \nonumber
\end{eqnarray}
\begin{eqnarray}
J_{3_{\lambda_4}}=K_{3_{\lambda_4}}= -\frac{2p_z(E+m-p_y)}{m(E+m)},
\end{eqnarray}
belonging, thus, to the Lounesto class-6, standing for a dipole spinor. A straightforward examination of all bilinear forms introduced above, shows that Fierz-Pauli-Kofink identities are automatically reached: $J_{\mu}J^{\mu}=0$, $K_{\mu}K^{\mu}=0$, both clearly being (null) invariants, moreover, we also have $J_{\mu}K^{\mu}=0$ and $J_{\mu}K_{\nu}-K_{\mu}J_{\nu} =0$. Notice the following $K_{\mu}=-J_{\mu}$ for $\lambda_1(\textbf{p})$ and $\lambda_2(\textbf{p})$ spinors and $K_{\mu}=J_{\mu}$ for $\lambda_3(\textbf{p})$ and $\lambda_4(\textbf{p})$ spinors, such results are consistent with expectations of Lounesto \cite{lounestolivro}.

Under the action of the charge-conjugation operator, the Ahluwalia class-7 spinors behave like
\begin{eqnarray}
&&\mathcal{C}\lambda_1(\textbf{p})=-\lambda_4(\textbf{p}), \quad \mathcal{C}\lambda_2(\textbf{p})=\lambda_3(\textbf{p}), 
\\
&&\mathcal{C}\lambda_4(\textbf{p})=-\lambda_1(\textbf{p}), \quad \mathcal{C}\lambda_3(\textbf{p})=\lambda_2(\textbf{p}),
\end{eqnarray}
in which $\mathcal{C}$ stands for the charge-conjugation operator. The right observance of the last set of equations evinces that the charge-conjugation operator plugs the $\lambda$'s particle-antiparticle counterparts.
Thus, Ahluwalia class-7 spinors --- up to a constant multiplicative factor and an appropriate dual re-definition, it may lead to a more involved physical framework--- being expansion coefficient functions of a quantum field, bringing to the light a new set of dark fields.
   
Notice because $\gamma_5$ commutes with $\Gamma_7$, the Ahluwalia class-7 spinors  also becomes eigenspinors of the chirality operator, $-i\gamma_{0123}\equiv\gamma_{5}$, furnishing the relations
\begin{eqnarray}
&&-i\gamma_{0123}\lambda_j = -\lambda_j, \quad j=1, 2 \label{chiral1}
\\
&& -i\gamma_{0123}\lambda_j = +\lambda_j, \quad j=3, 4. \label{chiral2}
\end{eqnarray}

Interesting enough, accordingly to Ref \cite{benn1987} and later verified in \cite{crumeyrolle2013}, driven by an exhaustive analysis of irreducible representations and the Clifford (sub)algebra, it is possible to write a Dirac spinor ($\psi_D$) from dipole spinors
\begin{eqnarray}
\psi_{D} = \lambda_{+} + \lambda_{-},
\end{eqnarray}
namely \emph{Weyl condition}. The sub indexes refers to the behaviour of the $\lambda$ spinors under action of the chiral operator in \eqref{chiral1} and \eqref{chiral2}. A straightforward calculation, shows up two possibilities to define Dirac spinor 
\begin{eqnarray}
&&\psi_{D_{1}}(\textbf{p}) = \lambda_{1}(\textbf{p}) + \lambda_{3}(\textbf{p}),
\\
&&\psi_{D_{2}}(\textbf{p}) = \lambda_{2}(\textbf{p}) + \lambda_{4}(\textbf{p}),
\end{eqnarray}
the above set of Dirac spinors belong to class-2 within Lounesto's classification. Moreover, from dipole spinors it is also possible to define another interesting set, standing for the following
\begin{eqnarray}
&&\Psi_{1}(\textbf{p}) = \lambda_2(\textbf{p}) + \lambda_3(\textbf{p}),\label{dual1}
\\
&&\Psi_{2}(\textbf{p}) = \lambda_1(\textbf{p}) + \lambda_4(\textbf{p}). \label{dual2}
\end{eqnarray}
The above spinors holds the following properties $\mathcal{C}\Psi_{1}(\textbf{p})=+\Psi_{1}(\textbf{p})$ and $\mathcal{C}\Psi_{2}(\textbf{p})=-\Psi_{2}(\textbf{p})$.

\section{Concluding remarks }\label{remarks}
We finalize by elucidating how useful tool shown to be the roots of the identity matrix in the $(1/2,0)\oplus(0,1/2)$ representation space. Given machinery allow one to define a set of four spin one-half orthonormal spinors and the associated quantum field --- such a mechanism reveals new classes of fermions and also bosons. Possibly bringing to light new candidates to compose what we call ``\emph{Beyond the Standard Model}''. 

The focus of this work are the \emph{Ahluwalia class-7 spinors}, which, as can be seen, are eigenspinors of the chirality operator. The aforementioned spinors enlarge our comprehension about the Lounesto's class-6, in which the main characteristic lies in the dark feature, standing for a entirely new set of mass dimension one matter field.


\bibliographystyle{unsrt}
\bibliography{refs}

\end{document}